
 {
\scrollmode
\NPrefs
\def\define#1#2\par{\def#1{\Ref#1{#2}\edef#1{\noexpand\refmark{#1}}}}
\def\con#1#2\noc{\let\?=\Ref\let\<=\refmark\let\Ref=\REFS
         \let\refmark=\undefined#1\let\Ref=\REFSCON#2
         \let\Ref=\?\let\refmark=\<\refsend}

\define\WITT
E.~Witten, Phys. Rev.{\bf D44} (1991) 314.

\define\AMS
G.~Mandal, A.~M.~Sengupta and S.~R.~Wadia, Mod. Phys. Lett.{\bf A6} (1991)
1685.

\define\BARS
I. Bars, USC- 91/HEP/B3 (1991);
I. Bars, S. Sfetsos, USC-91/HEP- B6 (1991).

\define\HORNE
J. Horne and G. Horowitz, Nucl. Phys.{\bf B368} (1992) 444.

\define\HORAVA
P. Horava, Phys. Lett.{\bf B278} (1992) 101.

\define\GIBBMAED
G. W. Gibbons and K. Maeda, Nucl. Phys,{\bf B298} (1988) 741;
D. Garfinkle, G. Horowitz and A. Strominger, Phys. Rev.{\bf D43} (1991) 3140;
C. G. Callan, R. C. Myers and M. J. Perry, Nucl Phys.{\bf B311} (1988) 673;.
G. Horowitz and A. Strominger, Nucl. Phys.{\bf B360} (1991) 197;
A. Shhapere, S. Trivedi and F. Wilczek, Mod. Phys. Lett.{\bf A6} (1991) 2677.

\define\HORNHORO
J.Horne and G. Horowitz, UCSBTH-92-11.

\define\SEN
A. Sen, TIFR/TH/92-20, April 1992.

\define\FAWAD
G. Veneziano, Phys. Lett.{\bf B265} (1991) 287; K. Meissner and G. Veneziano,
Phys. Lett.{\bf B267} (1991) 33; Mod. Phys. Lett.{\bf A6} (1991) 3397;
M. Gasperini, J. Maharana and G. Veneziano, Phys. Lett.{\bf 272} (1991)277;
A. Sen, Phys. Lett.{\bf B271} (1991) 295; Phys. Lett.{\bf B274} (1991) 34;
S. F. Hassan and A.Sen, Nucl. Phys.{\bf B375} (1992) 103.

\define\CHUNGTYE
S. Chung and S. H. Henry Tye, CLNS 91/1127, Jan. 1992.

\define\ALOK
S. Kar and A. Kumar, IP/BBSR/92-18, March 1992.

\define\DIJK
R. Dijkgraaf, H. Verlinde and E. Verlinde, PUPT-1252, IASSNS-HEP-91/22.

\define\RAITEN
E.~Raiten, Fermilab-PUB-91/338-T, December 1991.

\define\ISHIBASHI
N. Ishibashi, M. Li and A. R. Steif, Phys.~Rev.~Lett. {\bf 67} (1991) 3336.

\define\GILBERT
G. Gilbert, UMDEPP 92-035, August 1991; UMDEPP 92-110, November 1991.

\define\FUCITO
A. Carlini, F. Fucito and M. Martellini, ROM2F-92/18, April 1992.

\define\KKSEN
S. K. Kar, S. P. Khastagir and G. Sengupta, IP/BBSR/92-35, May 1992.

\def\cg{{\cal G}}
\def\cb{{\cal B}}
\def\cR{{\cal R}}
\def\d{\delta}
\def\pr{\partial}
\def\prb{\bar \partial}

\pubnum{92-28}\date{May, 1992}
\titlepage

\title{FOUR DIMENSIONAL 2-BRANE  SOLUTION IN CHIRAL GAUGED WESS-ZUMINO
-WITTEN MODEL}
\author{Swapna Mahapatra}\foot{e-mail address:
SWAPNA@TIFRVAX.BITNET}
\address{Tata Institute of Fundamental Research, Homi Bhabha Road, Bombay
400 005, India}
\bigskip
\abstract
An exact conformal field theory describing a four dimensional
2-brane solution is found by considering a chiral gauged Wess-Zumino
-Witten theory corresponding to $SL(2, R)\times R$, where one gauges
the one dimensional $U(1)$ subgroup together with a translation in
$R$. The backgrounds for string propagation are explicitly obtained
and the target space is shown to have a true curvature singularity.
\endpage
In recent times the study of exact solutions of string theory with
black hole behaviour has received much attention. A particularly
interesting solution of the graviton-dilaton field equations has
recently been found \WITT\AMS, where it was shown that the solution
exhibits all the characteristic features of the 2 dimensional black
hole solution. So the important question is how to generate these
backgrounds for string propagation and what happens to string
propagation in these singular backgrounds. The exact conformal field
theory description of these theories have been given by Witten \WITT,
where he has shown that the 2 dimensional black hole solution emerges
by considering a gauged Wess-Zumino-Witten (WZW)
model of $SL(2, R)/U(1)$. This novel way of generating 2d black hole
backgrounds by gauging a WZW model has also been used to the construct
higher dimensional analogs of this solution \BARS\HORNE\HORAVA.
The black hole solutions of the low energy effective
string theories have been analyzed by several people \GIBBMAED.
Rotating four dimensional charged black hole solutions have also been
obtained recently \HORNHORO\SEN, where the black hole carries mass,
axionic charge and angular momentum. In ref. \SEN\ a twisting procedure
\FAWAD\ has been used to obtain charged black hole solutions in heterotic
string theory by considering a more general action involving graviton,
dilaton and antisymmetric tensor gauge field.

In the first approach, where the black hole solution is obtained by
considering a gauged WZW model (where the gauge fields are dynamical),
one gauges an anomaly free subgroup of the group $G$. Depending on the
subgroup one gauges, one obtains either Euclidean or Minkowski black hole.
In the context of $SL(2, R)/U(1)$ gauged WZW model, two different gaugings
namely, axial and vector gauging had been considered previously. In a
recent interesting paper \CHUNGTYE, the authors have shown that there
exists another way to introduce gauge couplings into WZW theory which is also
anomaly free, hence the resulting theory is gauge invariant.
This is called chiral gauging where in the corresponding gauged WZW theory,
the gauge fields are chiral  meaning  each gauge field has only one
helicity and they belong to the subgroups $H_L$ and $H_R$ (which can
be different) of $G_L$ and $G_R$ respectively. The advantage of chiral
gauged WZW theories is that, it allows one to study the left right asymmetric
cases (for example heterotic string theories). Target space structure
of a chiral gauged WZW model has been analyzed recently in ref.\ALOK.

In this letter, we consider the $SL(2, R)$ WZW model with one free boson
(uncompactified) $X'$ added to it. This basically amounts to taking
$G = SL(2, R)\times R$. We then consider a chiral $U(1)$ gauging of $SL(2, R)$
along with a translation in $R$. We analyze the target space structure
of this model and find that the backgrounds corresponding to string propagation
are singular and they satisfy the field equations derived from the low
energy effective action. We specifically show the emergence of four
dimensional 2-brane
solution in the case under consideration and compare our solution with the
recent solution obtained in the context of perturbation of stringy
black hole \RAITEN, where a four dimensional black hole solution has been
obtained by adding two noncompact bosons ($X_1$ and $X_2$) to $SL(2, R)$
and then considering a vectorial gauging of the $SL(2, R)$ along with
translations in both $X_1$ and $X_2$. On the otherhand, if we consider a
compact boson, we generate a three dimensional electrically
charged black string
solution where we have a nontrivial gauge field background and an additional
scalar field along with
graviton and dilaton and antisymmetric tensor field background.

We now describe the conformal field theory construction yielding the
four dimensional 2-brane solution. The ungauged WZW action for
$SL(2, R)$ can be written as,
$$
S_{WZW} = {k\over 4 \pi} \int_{\Sigma}\,d^2z\,Tr\,(g^{-1}\,\pr g\,
g^{-1}\,\pr g) - {k\over 12 \pi}\int_{B}\,Tr\,(g^{-1}\,dg \wedge\,
g^{-1}\,dg \wedge\,g^{-1}\,dg), \eqn\one
$$
where $B$ is a three manifold with boundary $\Sigma$, $g$ is an element
of $SL(2, R)$ and $k$ is the level of the WZW model. When we consider the
chiral gauging, each gauge field has
only one type of chirality as discussed before and we have,
$$
A_{z}^L = A_{\bar z}^{R} = 0;
\qquad v^L = v^L(\bar z) , v^R = v^R(z), \eqn\two
$$
where $v_L$ and $v_R$ are the infinitesimal gauge transformation parameters.
Now the action corresponding to chiral gauged WZW model can be written as,
$$
S = S_{WZW} (g) + {k\over 2 \pi} \int\,d^2z\,Tr\,[A_{z}^R\,g^{-1}\,
\prb g + {A_{\bar z}^{L}}\,\pr g g^{-1} + A_{z}^R\,g^{-1}\,{A_{\bar z}^{L}}\,
g]. \eqn\three
$$
We introduce real co-ordinates $r$, ${\phi}_L$ and ${\phi_R}$ and parametrize
the group manifold by,
$$g = \exp({i\over 2}\phi_L\sigma_2)\,\exp({r\over 2}\sigma_1)\,
\exp({i\over 2}\phi_R\sigma_2)\eqn\four
$$
where ${\sigma}_i$ are the Pauli matrices and the co-ordinates range over
$0\leq r < \infty$, $0\leq {\phi}_L < 2 \pi$, $-2 \pi \leq
{\phi}_R < \pi$ (we use notations of ref. \DIJK).
Now we introduce the free boson $X'$ and couple the gauge
fields to it. The couplings can be introduced in two ways where the
gauged actions corresponding to the free boson is given by \ISHIBASHI,
$$
S_{FB} = {1\over\pi} \int\,d^2\,z\,(\pr X' + c_R\,
A_{z}^R)\,(\prb X' + c_L\,{A_{\bar z}^{L}}), \eqn\five
$$
and,
$$
S_{FB} = {1\over \pi} \int\,d^2\,z\,(\pr X' \prb X' +
c_L\,{A_{\bar z}^{L}}\,\pr X' + c_R\,A_{z}^R\,\prb X'). \eqn
\six
$$
where $\pr X'$ and $\prb X'$ are the chiral
currents and $c_L$ and $c_R$ are the gauge couplings. In the chiral gauged
model, we have the freedom of choosing them different also. The second action
eqn.\six\ is invariant under the world sheet gauge transformations,
$$\eqalign{\d X'&= 0,\cr
\d {A_{\bar z}^{L}}&= - {\prb v^{L}(\bar z)},\cr
\d {A_z}^R&= \pr v^{R}(z).\cr}\eqn\seven
$$
However, we choose to work in the first formalism, where the gauged action
is given by eqn.\five\ and we prefer to choose the gauge couplings to be equal,
namely $c_L = c_R = c$ though we have the option of taking them different.
In the context of two dimensional charged black hole solution of Ishibashi
et al \ISHIBASHI, they consider the second action so that it leads to a
blackhole with a real gauge field for all values of $c$. Since in our case,
we consider the uncompactified boson, we do not have the nontrivial
background gauge
fields and we can consistently work with the first gauged action. The
gauge transformations corresponding to the various fields are given by,
$$
\eqalign{\d g&=v_{\bar z}^{L} g - g v_{z}^{R},\cr
\d {A_{\bar z}^{L}}&=- {\bar \partial}v_{\bar z}^{L},\cr
\d A_{z}^R&=\pr v_{z}^{R} ,\cr
\d X'(z, \bar z)&=c(v_{\bar z}^{L} - v_{z}^{R}).\cr}\eqn\eight
$$
With this choice of gauge transformations, the total action $S_{WZW}$ +
$S_{gauged}$ is found to be gauge invariant, where
$$\eqalign{S_{gauged} &= {k\over 2 \pi} \int\,d^2\,z [A_{z}^R\, (\pr
{\phi}_R + \cosh r\,\pr {\phi}_L + {2 c\over k}\,\prb X') + {A_{\bar z}^{L}}\,
(\pr {\phi}_L + cosh r\,\pr {\phi}_R + {2 c\over k}\,\pr X') \cr
&- A_{\bar z}^{L}\,A_{z}^{R} (\cosh{r} - {2c^{2}\over k})]\cr}\eqn\nine
$$
and the ungauged WZW action is given by,
$$S_{WZW} = {k\over 4\pi}\int d^{2}z \left( \prb r\pr r -
\prb\phi_{L}\pr\phi_{L} - \prb\phi_{R}\pr\phi_{R}
-2\cosh{r}\prb\phi_{L}\pr\phi_{R}\right) + {1\over\pi}\int d^{2}z \pr
X'\prb X'\eqn\ten
$$
By using the equations of motion of the gauge fields, we now redefine the
gauge fields as follows:
$$\eqalign{A_{\bar z}^{L} &= \hat A_{\bar z}^{L} + {(\prb \phi_R +
\cosh r \prb\phi_L + (2 c/k) \prb X')\over\cosh r -
(2 c^2)/k} ;\cr
A_{z}^{R} &= \hat A_{z}^{R} + {(\pr \phi_L + \cosh r \pr\phi_R +
(2 c/k)\pr X')\over\cosh r - (2 c^2)/k}.\cr}\eqn\elvn
$$
using these redefinitions, the action can be written as,
$$\eqalign{S &= {k\over 4 \pi}\int d^2 z [\prb r \pr r +
(\prb\phi_L \pr\phi_L +
\prb \phi_R \pr \phi_R){(\cosh r + 2 c^2/k)\over\cosh r - 2 c^2/k} \cr &+
\pr X' \prb X' (4/k {\cosh r\over\cosh r - 2 c^2/k}) +
(\prb \phi_L \pr \phi_R + \pr \phi_L \prb \phi_R){(1 + 2c^2/k \cosh r)
\over\cosh r - 2 c^2/k} \cr &+ (\prb X' (\pr \phi_L + \pr \phi_R) +
\pr X'(\prb \phi_L + \prb \phi_R))
{2 c/k(1 + \cosh r)\over\cosh r - 2 c^2/k} \cr\!\!\!\!\!\! &+
(\prb \phi_L \pr \phi_R - \pr \phi_L \prb \phi_R){(2 c^2/k \cosh r - 1)
\over\cosh r - 2 c^2/k} \cr\!\!\!\!\!\! &+
(\prb X' (\pr \phi_L - \pr\phi_R) -
\pr X' (\prb \phi_L - \prb \phi_R)){2c/k(1 - \cosh r)
\over\cosh r - 2 c^2/k} \cr &-
\hat A_{\bar z}^{L}\,\hat A_{z}^{R}\,(\cosh r - 2 c^2/k)]\cr}\eqn\twlv$$
By comparing this with the $\sigma$-model action,
$$S = {1\over\pi} \int d^2 z [G_{\mu\nu} + B_{\mu\nu}] \pr X'^{\mu}
\prb X'^{\nu},\eqn\thirtn$$
where $\mu$ and $\nu$ run over the four co-ordinates $r$, $\phi_L$,
$\phi_R$ and $X'$, we obtain the target space structure of the
corresponding chiral gauged WZW theory. The background metric, antisymmetric
tensor field are given by
$$G =\pmatrix {-1&0&0&0\cr 0&-{(\cosh r + 2 c^2/k)\over{\cosh r - 2
c^2/k}}&-{(2 c^2/k \cosh r + 1)\over{\cosh r - 2 c^2/k}}&{-2 c/k(1 + \cosh r)
\over{\cosh r - 2c^2/k}}\cr 0&-{(2 c^2/k \cosh r + 1)\over{\cosh r - 2 c^2/k}}
&-{(\cosh r + 2 c^2/k)\over{\cosh r - 2 c^2/k}}&{-2 c/k(1 + \cosh r)\over
{\cosh r - 2 c^2/k}}\cr 0&{-2 c/k(1 + \cosh r)\over{\cosh r - 2 c^2/k}}&
{-2 c/k(1 + \cosh r)\over{\cosh r - 2 c^2/k}}&-{4/k\over{\cosh r - 2 c^2/k}}
\cr}\eqn\fortn$$
and
$$B =\pmatrix {0&0&0&0\cr 0&0&-{(2 c^2/k \cosh r - 1)\over{\cosh r - 2 c^2/k}}
&-{2 c/k(\cosh r - 1)\over{\cosh r - 2 c^2/k}}\cr 0&{(2 c^2/k \cosh r - 1)
\over{\cosh r - 2 c^2/k}}&0&-{2 c/k(1 - \cosh r)\over{\cosh r - 2 c^2/k}}\cr
0&{2 c/k(\cosh r - 1)\over{\cosh r - 2 c^2/k}}&{2 c/k(1 - \cosh r)\over
{\cosh r - 2 c^2/k}}&0\cr}\eqn\fiftn$$
By requiring that these  background fields must be an extremum of the
low energy string effective action, we find that the contribution to
the dilaton field is given by,
$$\phi = - \ln(\cosh r - 2 c^2/k) + {\rm constant}\eqn\sixtn$$
Now notice that the background values we have obtained, depend only on one
of the co-ordinate and are independent of three of the other co-ordinates.
This is very interesting as it is well known that if $G$ and $B$ are
only functions of time (in the present case it is $r$), their
invariance under general co-ordinate transformation and a specific
transformation of $B$ always allows to bring them into the form,
$$G=\pmatrix{-1&0\cr
0&\cg(r)\cr}; \qquad B=\pmatrix{0&0\cr
0&\cb(r)\cr}.\eqn\sevntn$$
Now we show that the background fields we have obtained, actually
satisfy the low energy field equations derived from the string effective
action. The field equations are

$$\eqalign{
{(\dot \Phi)}^2 - 1/4 Tr[(\cg^{-1} \dot\cg)^2] +
1/4 Tr[(\cg^{-1}\dot\cb)^2] - V &= 0, \cr
(\dot \Phi)^2 - 2(\ddot \Phi) + 1/4 Tr[(\cg^{-1} \dot\cg)^2]
- 1/4 Tr[(\cg^{-1} \dot\cb)^2] - V + {\pr V\over\pr\Phi} &= 0,\cr
- \dot\Phi\dot\cg + \ddot\cg - \dot \cg\,\cg^{-1}\,\dot\cg
- \dot\cb\,\cg^{-1}\,\dot \cb &= 0,\cr
- \dot\Phi\,\dot\cb + \ddot\cb - \dot\cb\,\cg^{-1}\,\dot\cg
- \dot\cg\,\cg^{-1}\,\dot\cb &= 0,\cr}\eqn\eightn
$$
where,
$$\Phi = \phi - \ln{(\sqrt{det\cg})}.\eqn\nintn$$
Here dots represent derivatives with respect to $r$. We found that the
above values for the background fields satisfy the field equations for a
constant potential $V = 1$. In general, $V$ can be a function of $\Phi$.

The curvature scalar $\cR$ is given by the expression,
$$\cR = 2(\pr_r^2 \ln{(\sqrt{det\cg})}) +
(\pr_r \ln{(\sqrt{det\cg})})^2 -
1/4 Tr[(\pr_r \cg)(\pr_r \cg^{-1})],\eqn\twenty$$
which gives
$$\cR = {[7 - 8 c^2/k(\cosh r - 2 c^2/k) - 28 c^4/{k^2}]\over
{2 (\cosh r - 2 c^2/k)^2}}\eqn\twnton$$
Now we want to make some suitable coordinate redefinitions in order to
bring the background fields and the curvature to simpler forms. First of
all we define,

$$\eqalign{\cosh r - 2 c^2/k &= 2 \hat r ;\cr
\lambda &= 2 c^2/k;\cr
{X\over\sqrt{1 + \lambda}} &= 1/2(\phi_L + \phi_R);\cr
{i\,t\over\sqrt{1 - \lambda}} &= 1/2(\phi_L - \phi_R);\cr
{2\over\sqrt k}X' &= \tilde {X'}.\cr}\eqn\twntwo$$
The line element is now given by,
$$\eqalign{
ds^2 &=[(1 + {1 + \lambda\over 2\hat r})\, dX^2 + (1 + {\lambda
\over 2\hat r})\, d\tilde {X'}^2 - (1 - {1 - \lambda\over 2\hat r})\, dt^2
\cr\!\!\!\!\!\!\!\!\!\! &+ \sqrt{\lambda\over{1 + \lambda}}\,
(1 + {1 + \lambda\over 2\hat r})\,(2dX\,d\tilde X')
+ (1 - {1 - \lambda\over 2\hat r})^{-1}\,
(1 + {1 + \lambda\over 2 \hat r})^{-1}\,{d\hat r^2\over\hat
r^2}]\cr}\eqn\twnthr$$
We have dropped an overall $k/4$ term. At this point we would like
to compare our solution with the four dimensional solution obtained by
adding two free noncompact bosons to $SL(2, R)$ \RAITEN.
Comparing the $d{\hat r}^2/{\hat r}^2$ terms in the two cases, we get
the constraint,
$$
{1 - \lambda\over 2} = (1 + \lambda_1 + \lambda_2)\eqn\twnfor
$$
where ${\lambda}_1$ and ${\lambda}_2$ are the gauge couplings corresponding
to the two bosons. We make further co-ordinate redefinitions such as,
$$\eqalign{X &= - \sqrt{1 + \lambda}\,x_2 \cr
\tilde X' &= x_1 + \sqrt{\lambda}\, x_2}\eqn\twntfiv$$
With this choice, the expressions for the metric and the antisymmetric
tensor field actually simplify a lot even though the metric is not
completely diagonal, namely,
$$\eqalign{ds^2 &= (1 + {\lambda\over 2\hat r})\,d{x_1}^2 +
(1 + {1\over 2\hat r})\,d{x_2}^2 - (1 - {1 - \lambda\over
2\hat r})\,dy^2 \cr \!\!\!\!\!\! &+ {d\hat r^2\over\hat r^2}\,(1 -
{1 - \lambda\over 2\hat r})^{-1}\,(1 + {1 + \lambda\over
2\hat r})^{-1} - {\sqrt{\lambda}\over 2\hat r}\,
(dx_1\,dx_2 + dx_2\,dx_1)\cr}\eqn\twntsx$$
and,
$$B_{t x_1} = \sqrt{-\lambda\over {1 - \lambda}}\,[1 - {1 - \lambda\over
2 \hat r}]; \qquad
B_{t x_2} = \sqrt{-1\over{1 - \lambda}}\,[-{(1 - \lambda)\over 2\hat r}]
\eqn\twntsvn$$
Notice that eqn.\twntsx\ tells us that our metric is asymptotically flat.
The cross term precisely goes to zero as $\hat r \rightarrow \infty$. Now
if we compare eqn.(2.15) of \RAITEN, we find that for ${\lambda}_1
= -{\lambda\over 2}$ and ${\lambda}_2 = {-1\over 2}$, the four dimensional
solution obtained there, exactly matches with ours. Note that, these values
of ${\lambda}_1$ and ${\lambda}_2$ are also consistent with our earlier
eqn.\twnfor. Further, we compare eqn.\twntsvn with eqn.(2.16) of ref. \RAITEN\
for the value of the antisymmetric tensor field and we find that both
the values of $B_{t x_1}$ exactly match with each other and the values
of $B_{t x_2}$ differ by a constant, which is perfectly fine as we have the
flexibility of adding constants to $B$ fields which does not change the
field strength $H = d\,B$. The curvature with the redefined co-ordinates
is given by,

$$\cR = {7 - 8 \lambda\hat r - 7 {\lambda}^2\over 8\hat r^2}\eqn\twntet$$
This shows that even if the metric components are ill defined at $2\hat r
= 0$, $\lambda$ and $1 + \lambda$, the genuine curvature singularity
is only at $\hat r = 0$. Also as $\hat r \rightarrow \infty$,
curvature goes to zero confirming the fact that the metric is asymptotically
flat.

There are several interesting comments in order. First of all, we notice
from eqns. (13) and (14) that as $k \rightarrow \infty$, our solution
reduces to that of the three dimensional solution of chiral gauged WZW
model \ALOK\ and equivalently to the black string solution of
Horne and Horowitz \HORNE\
with proper identifications. Also in the limit $c \rightarrow 0$,
{\it i.e.} when $R$ is not gauged, one gets the three dimensional
black string \HORNE\ and a flat(noncompact) direction. We have commented
in the beginning on adding compact boson to $SL(2, R)$, where we
generate a three dimensional electrically charged black string solution.
To summarize,
we have described here the string propagation in singular backgrounds by
considering a chiral gauged WZW model and we have obtained explicit
expression for the four dimensional 2-brane solution. The WZW approach
plays a crucial role in identifying the background fields. We have also
shown that our metric is asymptotically flat and the target space has a true
curvature singularity at $\hat r = 0$. We have considered various limits
of our solution to show that it reduces to other known solutions. It will
be interesting to study the stability under perturbations of this
solution \GILBERT\ (also see refs.\FUCITO\RAITEN\ for related work).
Further analysis of these theories will hopefully shed some
light on the understanding of string theory and black hole.

\noindent{\bf{Acnowledgements}}: I would like to thank S. Fawad Hassan and
Ashok
   e
Sen for illuminating discussions.

\noindent{Notes added:
After this work was completed, I became aware of a preprint \KKSEN\
where similar results have been obtained}.
\refout
\bye